# Powder reuse cycles in electron beam powder bed fusion – Variation of powder characteristics

Gitanjali Shanbhag[1], Mihaela Vlasea[1]

[1]University of Waterloo, Waterloo, ON N2L 3G1, CANADA

**Abstract:**

A path to lowering the economic barrier associated with the high cost of metal additively manufactured components is to reduce the waste via powder reuse (powder cycled back into the process) and recycling (powder chemically, physically, or thermally processed to recover the original properties) strategies. In electron beam powder bed fusion, there is a possibility of reusing 95 - 98% of the powder that is not melted. However, there is a lack of systematic studies focusing on quantifying the variation of powder properties induced by number of reuse cycles. This work compares the influence of multiple reuse cycles, as well as powder blends created from reused powder, on various powder characteristics such as the morphology, size distribution, flow properties, packing properties and chemical composition (oxygen and nitrogen content). It was found that there is an increase in measured response in powder size distribution, tapped density, Hausner ratio, Carr index, basic flow energy and specific energy, dynamic angle of repose, oxygen, and nitrogen content, while the bulk density remained largely unchanged.

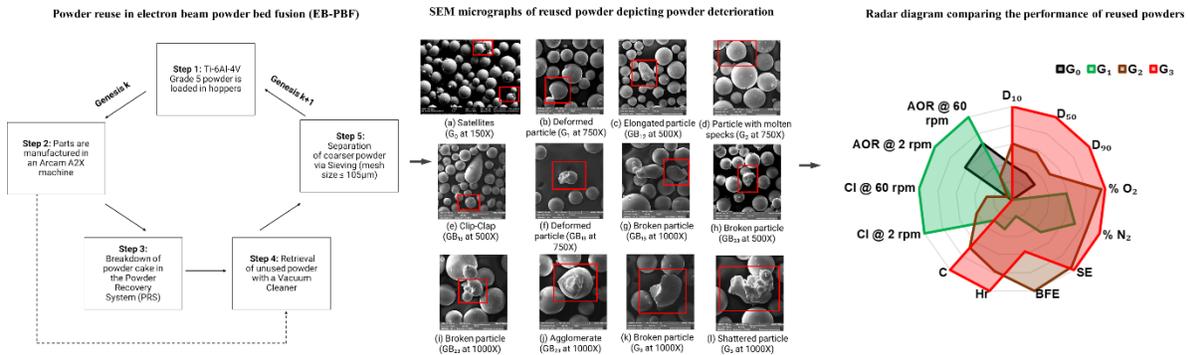

Figure: Graphical abstract

# 1 Introduction

The performance of a powder-based additively manufactured part highly relies on the quality of the powder feedstock properties. Vock et al. [1] stated that the correlation between bulk powder behavior, powder layer organization and final part quality is still not well understood. To improve the understanding of the powder-process-part relationship, there is a need to delve into quantifying the variability in powder properties, either as a function of print cycles, handling, or batch-to-batch variability and investigate their effects on the resulting part properties.

## 1.1 Powder parameters of interest for electron beam powder bed fusion (EB-PBF) processes

When looking at powder feedstock for powder bed fusion (PBF) processes, Popov et al. [2] identified that some of the fundamental powder characteristics that need to be assessed include shape, particle size, composition, gas infusions, flowability, tendency to oxidize and sintering/melting conditions.

Particle morphology has a considerable influence on the powder bed packing density, and consequently on the final component density, where the more irregular the particles, the lower the final density [3]. Higher apparent densities are preferred, as they provide better heat conduction, reducing the risk of sample swelling and overheating [4] in the EB-PBF process. In terms of powder morphology, spherical or regular, equiaxed particles, are less cohesive and tend to flow freely, arrange and pack more efficiently than irregular or angular particles [5]. As shape deviates from spherical, the interparticle friction increases and furthermore detrimentally affects the powder flowability and packing efficiency. Powder morphology can also affect the mechanical properties, microstructure, and surface finish of the final component. Consequently, Medina [6] emphasizes that powder morphology examination should be performed to identify particle shape, the presence of satellites, foreign particles or contamination.

Most powders for PBF processes are manufactured via atomization. Powders used in the EB-PBF process are typically manufactured via plasma atomization (PA). PA is the process of melting a wire spool feedstock of metal with a plasma torch and cooling it in an inert tower [7]. Powder particles manufactured by PA are usually spherical with minimal satellites and pores. It is of interest to quantify the effect of powder morphology on the EB-PBF processes as a function of powder reuse.

Particle Size Distribution (PSD) is a significant parameter in determining the minimum layer thickness, the minimum achievable feature size in final parts; and affects the powder-energy source interaction. The EB-PBF process uses a nominal PSD between 45 μm to 105 μm. Simchi [8] explained that a deviation in the PSD can lead to in-situ powder segregation and layer streaking due to coarser particles being pushed away from the powder bed. This could lead to variations in build quality. Consequently, it is of interest to quantify the effect of PSD on the EB-PBF process as a function of powder reuse.

The flowability of the powder is also very important, to ensure uniform layers when dispensed, distributed, and/or spread onto the build area. It is generally understood that in order to obtain powder layers with homogeneous density, it is important to ensure that the powders are free-flowing and exhibit good flow properties. Powder flowability behavior can be correlated to the size, shape, moisture content, and packing efficiency of the powder particles [9]. For example, larger and spherical particles tend to flow better than smaller and irregular particles. The Angle of Repose (AOR) can be used to characterize the flowability of powders [1]. The AOR is affected by various cohesive forces: Van der Waals, electrostatic and capillary, as well as the contact forces between powder particles. Teferra [10] states that powders that show a low AOR are categorized as non-cohesive, highly flowable powders and can be transported using gravitational force or very little energy. Powders with high AOR values are characteristic of cohesive powders and may lead to sporadic or intermittent flow. Powder rheometry characterization provides a suite of in-depth powder performance metrics such as tap density, apparent density, dynamic flow testing, dynamic angle of repose, shear index and cohesiveness. It is of interest to quantify the effect of powder flowability on the EB-PBF processes as a function of powder reuse.

Inert gas fusion analysis (i.e., LECO) provides quantitative data on the absolute oxygen and nitrogen content in the powders. This analysis is essential to understand whether the evolution of the oxygen and nitrogen content in the reused Ti-6Al-4V powder, is within the allowable concentration limits defined by ASTM 2924-14 [11] ( i.e. < 0.20 wt.% for oxygen and < 0.05 wt.% for nitrogen). It is of interest to quantify the effect of oxygen and nitrogen content on the EB-PBF processes as a function of powder reuse.

## 1.2  Powder parameter studies in EB-PBF of Ti-6Al-4V

There has been an increasing interest in fabricating parts using EB-PBF. The EB-PBF process offers distinct advantages when compared to other metal PBF processes. In EB-PBF, components are manufactured under a controlled vacuum and at an elevated temperature [12]. This environment is beneficial as it has the potential of having minimal impact on the chemical composition of the powders and is suitable for reactive materials such as titanium alloys [13] and it results in lower residual stresses in the deposited material by virtue of the elevated process temperature [14]. In addition, the EB-PBF process enables faster build rates owing to the high power and fast beam scanning [4]. It is also a "tool-less" fabrication technology which requires neither fixtures nor tooling to obtain a custom part geometry, within tolerance limits. EB-PBF technology also permits blends of used and new powders to be recycled back into the system, therefore this technology has the potential to be an environmentally friendly process.

To continually improve the process, it is important to identify, address, and overcome some of the process limitations. Debroy et al. [15] have identified that the cost of a manufactured part, in PBF processes, is essentially contributed by the additive manufacturing equipment, feedstock material, manufacturing and indirect costs [16]. Specifically, when looking at the cost of 1 build via EB-PBF, Baumers et al. [17] have identified that the feedstock powder makes up to ~ 28% of the total cost of the build. Therefore, we can conclude that the cost of the EB-PBF process heavily relies on the reusability of powders and may not be cost-affordable for complex applications if the un-melted powder in the build bed is not reused. However, there has been limited published information available on the role of feedstock powder on the final material properties. Thus, the evaluation of the maximum number of allowable powder reuse cycles is an essential factor to assess process affordability for a specific part or application. Reusing Ti-6Al-4V powder in EB-PBF can result in changes in the chemical composition, powder morphology, powder size distribution, and flowability, resulting in changes in the properties of the solidified material.

Powder morphology has been observed to be modified after reusing Ti-6Al-4V powders in EB-PBF. Tang et al. [18] observed that the particles became less spherical, had fractures, protrusions, and concave sites after reusing. They also observed an increase in surface roughness and distortions in the final part. Strondl et al. [5] observed irregularities, impact marks, satellites, and stacked particles in the recycled powder. Similar findings were reported by Mohammadhosseini et al. [19], where satellites and aggregation of particles in the reused powder were observed.

Some studies also reported flowability results for virgin and reused powders. Tang et al. [18] observed that the reused powder showed lower flowability when measured by Hall flowmeter and attributed this to the blasting process, which lead to irregular particle morphology and impact marks on the particle surface. Contradictory to this, Mohammadhosseini et al. [19] observed no change in the flowability when measured by Hall flowmeter and Carney funnel after reuse.

Studies on the effect of reusing powder on chemical composition showed that reusing Ti-6Al-4V can lead to an increase in oxygen content. Sun et al. [20] observed that after 30 reuses, there was a 35% increase in oxygen content in the reused powder, from 0.15 wt. % in virgin powder to 0.20 wt.%. Petrovic and Niñerola [21] observed that the oxygen content exceeded the 0.20 wt.% limit after 12 reuse cycles, where the initial oxygen content in the virgin powder was 0.14 wt.%. They attributed this increase to the humidity pickup from the inner walls of the machine. Similarly, Tang et al. [18] observed that the oxygen content increased from 0.08 wt.% to 0.19 wt.% after 21 reuse cycles. They attributed this oxygen increase to exposure of powder to the air, including processing in the powder recovery system and sieving. An increase in oxygen content typically results in an increase in the strength but lowers the toughness and ductility of the final part [22]. Therefore, the mechanical properties of the parts will be varying with the number of reuse cycles.

Based on these studies, a powder suitability criteria can be created for reused powders (see Table 1), for use in the EB-PBF process, to understand what type of powder performance metric behavior is suitable with respect to the morphology, size distribution, sphericity, basic flow energy, specific energy, density, Hausner ratio, Carr index, cohesive index, angle of repose, oxygen content, and nitrogen content.

EB-PBF machines store about 100 – 180 kgs of powder (depending on the machine model) in the hoppers and it is often impossible to refill these hoppers with powder from the same reuse cycle. Hence blends of powders, either a mixture of virgin and reused or a mixture of different reuse cycles, are frequently used for manufacturing parts and such blend performances also need to be evaluated. These practices often pose challenges in isolating the effects of powder reuse in the above-mentioned studies. In addition, most published studies focus on assessing the effect of powder reuse on only a few specific powder performance metrics. Therefore, there is a need to assess the different powder characterization techniques and obtain the various powder performance metrics associated with these techniques to perform comprehensive powder reuse studies. This needs to be done by precisely isolating the powder blends used in the build, as well as performing a comprehensive study on effects of powder reuse on powder properties.

Table 1 Powder suitability criteria for use in the EB-PBF process

| Powder characteristic | Requirements | What other characteristics does this property have an influence on | Performance metrics that can help assess the powder property | Should this performance metric be maximized (↑), minimized (↓) or kept constant (↔) |
|---|---|---|---|---|
| Morphology | Spherical and equiaxed to increase flowability and avoid interparticle friction and mechanical interlocking | Powder bed packing density and the final component density, surface finish | Sphericity | ↑ |
| Apparent density | Should be high for improved heat conduction in the EB-PBF process | Heat conduction, Sample swelling and Overheating | $\rho_0$ | ↑ |
| Compressibility | High compressibility is desirable to be able to achieve higher packing density in the powder bed | Powder bed packing density, layer thickness, heat conduction | Carr Index ($C$) | ↓ |
| Particle Size distribution (PSD) | Stay within the manufacturer's size range and not drastically increase in order to obtain small feature sizes and thinner powder layers | Minimum layer thickness, Minimum achievable feature size, build quality, surface finish | $D_{10}$, $D_{50}$ and $D_{90}$ values | ↔ |
| Chemical composition | The oxygen and nitrogen concentration should stay within allowable concentration limits | Mechanical properties such as toughness, ductility of final parts, embrittlement | Oxygen (in wt.%) and nitrogen (in wt.%) | ↓ |
| Spreadability | A dynamic cohesive index closer to zero corresponds to a non-cohesive powder. A cohesive powder leads to a sporadic or intermittent flow while a non-cohesive powder leads to a regular flow | Powder bed packing density, powder layer uniformity, rake-ability, easy flow in the hoppers (i.e., powder storage). Decreased spreadability and flowability may contribute to more defects in the final part | Cohesive Index (CI) | ↓ |
| Flowability | Highly flowable powder minimizes the risk of mechanical interlocking and friction between particles and allows for smooth operation while raking and for uniform and homogenous distribution over the build plate | | Hausner ratio ($H_r$) | ↓ |
| | | | Basic Flow energy (BFE) | ↓ |
| | | | Specific Energy (SE) | ↓ |
| | | | Angle of repose (AOR) | ↓ |

The current comprehensive manuscript aims at advancing the preliminary study by the authors in Shanbhag and Vlasea [23], by identifying some of the major studies in literature and comparison of the authors results with those studies, determining a powder suitability criteria, describing the challenges related to a unified powder suitability criteria, analyzing robust trends with increase in powder reuse cycles for all characterization results, describing the potential impact of powder reuse on part properties as a precursor to a detailed follow-up study, establishing correlations between various powder performance metrics, evaluating relative performance of powder generations using a normalized radar diagram and formulating a unified powder suitability score that can be used by other researchers to compare powder performance.

## 2 Materials and Methods

The Ti-6Al-4V powder used was supplied by Advanced Powders & Coatings, Canada. The Grade 5 plasma-atomized powder (Batch number: P1321) was obtained in its pre-alloyed form with a size range of 45–105 μm. The chemical composition of the powder conforms to ASTM F2924 for a Grade 5 Ti-6Al-4V powder (according to batch information provided by powder supplier). A total of seven powder types were assessed for this study. The powder obtained from the supplier is referred to as the virgin powder (henceforth known as Genesis 0 or $G_0$[1]). Powder that was used once (Genesis 1 or $G_1$), was passed through a vibrating sieve (mesh size -140+325, i.e., 44 µm to 105 µm) to recover the powder for reuse. Genesis 2 ($G_2$) and Genesis 3 ($G_3$) powders were similarly obtained after printing parts with $G_1$ and $G_2$ powder, respectively. For every powder type, approximately 500 grams of powder was collected in a metal can. These cans were subsequently sealed and rolled/tumbled, on a jar-mill (Labmill 8000, Gardco, USA), at 40 revolutions per minute (RPM), for 24 hours in order to homogenize the sample before conducting any characterization. $G_1$ and $G_2$ powders were blended in equal wt.% to obtain the $GB_{12}$ blend and compare its properties with the other powder types. Similarly, $GB_{13}$ and $GB_{23}$ blends were obtained based on equal wt.% of powder constituents. The individual powder genesis were first collected in metal cans and rolled for 24 hours as mentioned earlier. These were then mixed in equal parts to create the respective blends. This blended powder was also collected in metal cans, sealed, and rolled for another 24 hours to ensure homogeneity and proper mixing in the blend. Therefore, the blended powders see a total rolling/tumbling time of 48 hours as compared to 24 hours for the unblended powders. All printing was done on an Arcam A2X machine using the default parameters provided for Ti-6Al-4V (Theme 5.2.52, Arcam A2X, GE Additive). To ensure that a consistent beam scanning strategy was used for all builds, the parts built were constant across the 3 printing cycles.

In order to assess the powder particle morphology, field emission microscopy (SEM, Zeiss Ultra & Tescan VEGA3) observations were performed. A Camsizer X2 (Retsch GmbH, Germany) was used to measure the PSD of the various powder types. The system uses the principle of digital image processing [24], where the dispersed particles pass in front of LED light sources and their shadows are captured with two digital cameras. The Retsch software (Retsch GmbH, Germany) analyzes the size of each particle captured by the camera and calculates the respective distribution curves.

---

[1] It should be noted that $G_0$ powder did not undergo any processing in the EB-PBF machine, blasting in the powder recovery system and sieving to remove the fine powder particles. Therefore, this powder type may be considered an anomaly for the various powder characterization. Nevertheless, the results for $G_0$ are presented in order to compare the results for all other powder types and use $G_0$ as the baseline.

Powder rheology was investigated using a powder rheometer (FT4, Freeman Technology, UK), rotating drum (Granudrum; Granutools, Belgium) and an automated tap density instrument (GranuPack; Granutools, Belgium). Performance metrics were collected, such as the basic flow energy (BFE), specific energy (SE), bulk density ($\rho_0$), tap density after 500 taps ($\rho_{500}$), Hausner ratio ($H_r$) and Carr index ($C$). To characterize the resistance to flow, eleven test cycles were run with a condition cycle run between each test. During the tests, the precision blade was rotated downwards and upwards through the fixed volume of powder to establish a flow pattern, where the resistance of the powder to the blade yielded the bulk flow properties. The BFE is defined as the energy required to displace a powder when the blade is moving downward, and the powder is constrained. As described by Freeman and Fu [25], the SE measures the powder's flowability as the blade rotates upward and the powder is unconfined as there is no enclosure at the top of the vessel. The GranuPack measures the evolution of the powder density as a function of the tap number to obtain a compaction curve (as shown in Figure 1), which is used to calculate the $H_r$ and $C$ values. $H_r$ is a number that is correlated to the flowability of a powder and is calculated using the formula $H_r = \rho_{500} / \rho_0$ where $\rho_{500}$ is the tapped density of the powder after 500 taps and $\rho_0$ is the initial bulk density of the powder. $C$ is related to the compressibility of a powder and is calculated by the formula $100\ (\rho_{500} - \rho_0)/ \rho_0$, where $\rho_0$ is the initial bulk density of the powder and $\rho_{500}$ is the final tapped density of the powder after 500 taps.

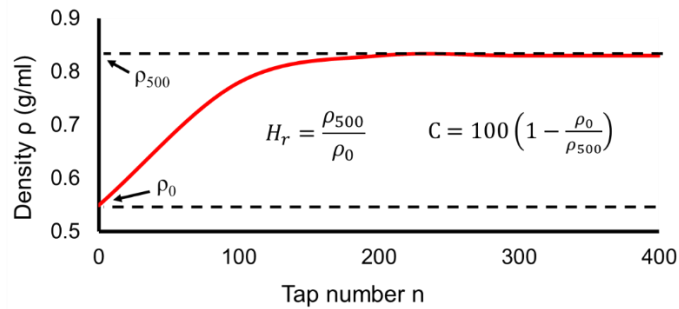

Figure 1 Example of compaction curve obtained by GranuPack - illustrating bulk density ($\rho_0$), tap density ($\rho_{500}$), Hausner ratio ($H_r$) and Carr index ($C$)

The GranuDrum instrument (Granutools, Belgium) is used to determine the dynamic angle of repose and the cohesive index. The GranuDrum is composed of a horizontal drum half-filled with powder that rotates around its axis at an angular velocity ranging from 2 rpm to 60 rpm. In total 17 velocities were tested, from 2 to 20 rpm at increments of 2 rpm followed by 25 to 60 rpm at increments of 5 rpm. To minimize internal sequence effects, a different velocity sequence was used for each replicate (3 replicates in total) such that the sequences for all replicates are minimally correlated. A camera takes snapshots for each angular velocity and the software calculates the flowing angle of powder (or AOR) and the cohesive index (CI) values.

Chemical analysis for the various powders were performed using inert gas fusion using a LECO TCH 600 (Leco Corporation, USA) instrument to analyze the oxygen and nitrogen content in the powders. All powder characterisation experiments were performed in triplicate and the average values are reported in this manuscript.

# 3 Results and Discussion

As the number of reuse cycles increases, properties such as chemical composition, surface features (e.g., surface roughness and overall particle roundness), and physical and thermal properties are expected to change. Therefore, understanding the powder behavior with reuse is important for both cost and quality control.

## 3.1 Observations of Changes in Powder Properties with Reuse Cycles

The powder is predominantly spherical in its as-received (or virgin) condition. Figure 2 presents the high magnification SEM micrographs for all different powder types. These micrographs help define and depict defects such as satellites (Figure 2(a)), elongated particles (Figure 2(c)), broken particles (Figure 2(g),(h),(i) and (k)), deformed particles (Figure 2(b) and (f)), particle with molten specks (Figure 2(d)), clip-clap (Figure 2(e)), shattered particles (Figure 2(l)), and agglomerates (Figure 2(j)). The nomenclature and morphology of $G_0$, $G_1$, $G_2$ and $GB_{12}$ powder types has been previously described in [23], [26]. Other authors have also reported defects such as non-spherical particles and presence of agglomerates after reusing. Sun et al. [20] observed noticeable deformations and distortions on the surface of reused EB-PBF powder particles. Cordova et al. [27] observed that reused powder, in laser PBF processes, exhibits a deformation towards a teardrop shape and a rougher surface due to remelting.

The deformed, broken, clip-clap and shattered particle defects are attributed to the recovery via the blasting procedure. The homogenization of powder via tumbling on a jar-mill, may result in numberless collisions between particles, and in friction and wear in presence of air and therefore the tumbling procedure may be another reason for these defects. The particles with molten specks and elongated particles are attributed to the temperature conditions that lead to overheating and smelting of the particles and satellites [26]. The agglomeration of powders is attributed to the high temperature of the process. Agglomerated particles result from the diffusion bonding obtained during preheating (to allow charge dissipation through the powder bed and reduce particle ejections resulting from the interaction of the electron beam during melting. The SEM micrographs (Figure 2) make it qualitatively evident that recovering and reusing the powder from the powder cake has changed the powder morphology significantly.

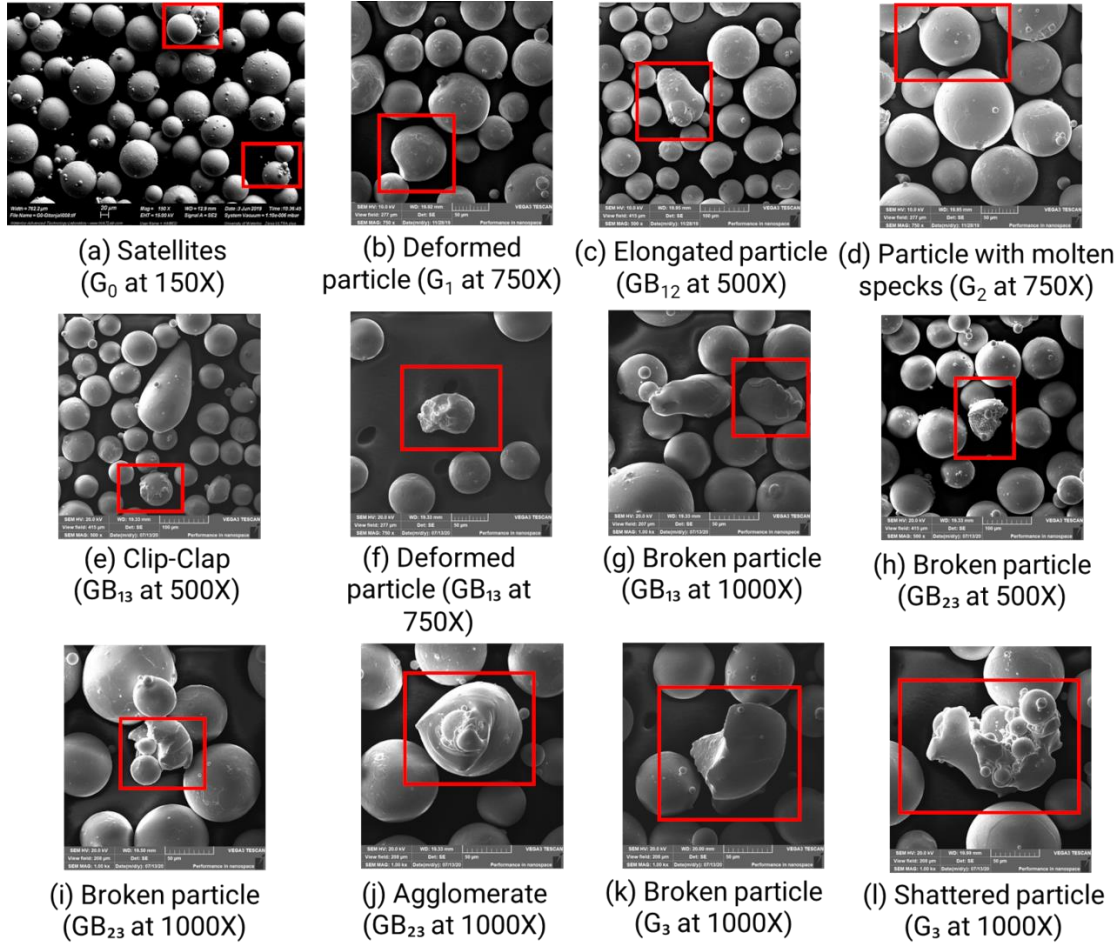

Figure 2 SEM micrographs for all powder types depict (a) Satellites in $G_0$ powder; (b), (f) Deformed particle in $G_1$ and $GB_{13}$ powders respectively; (c) Elongated particle in $GB_{12}$ powder; (d) Particle with molten specks in $G_2$; (e) Clip-Clap in $GB_{13}$ powder; (g), (h), (i), (k) Broken particles in $GB_{13}$, $GB_{23}$ and $G_3$ powders; (j) Agglomerate in $GB_{23}$ powder and (l) Shattered particle in $G_3$ powder. Nomenclature used to describe these micrographs was first defined by Popov et al. [26]. A Zeiss Ultra SEM instrument was used to capture the image shown in (a); all other images were captured using the Tescan VEGA3 instrument.

Figure 3 shows the $D_{10}$, $D_{50}$ and $D_{90}$ values of the different powders investigated. The $D_{50}$ (median value), is described as the diameter where half of the population lies below this value. Similarly, 90% and 10% of the distribution lies below the $D_9$ and $D_{10}$, respectively. It is worth noting that the $G_0$ powder has not undergone any processing in the EB-PBF machine, nor blasting or recovery through the sieve. This contributes to the discrepancies below 44 µm in the PSD of $G_0$ when compared to the other powder types. As expected, the $D_{10}$, $D_{50}$, $D_{90}$ values for $GB_{12}$ $GB_{13}$, and $GB_{23}$ lie between their respective genesis powders. This is because the authors ensured that the powders blends were made from equal wt.% of powder constituents and mixed thoroughly before characterization. The $D_{10}$, $D_{50}$, $D_{90}$ values for the individual genesis powders (i.e., $G_1$, $G_2$, $G_3$) show an increasing trend (Figure 3). In other words, we can say that the $D_{10}$, $D_{50}$, $D_{90}$ values increase with an increase in number of reuse cycles.

Slotwinski et al. [28] also reported an increase in particle size with increasing number of reuse cycles in LPBF. They associated this observation to the consolidation and loss of the small particles. Grainger [29] also noticed a disappearance of smaller particles with an increase in number of powder reuse cycles in LPBF. Although the LPBF process does not use the same energy source, nor result in a powder cake after sintering, the powders are exposed to sputter and undergo a sieving process to recover the powders for reuse; similarities in trends with EB-PBF are observed in the present work.

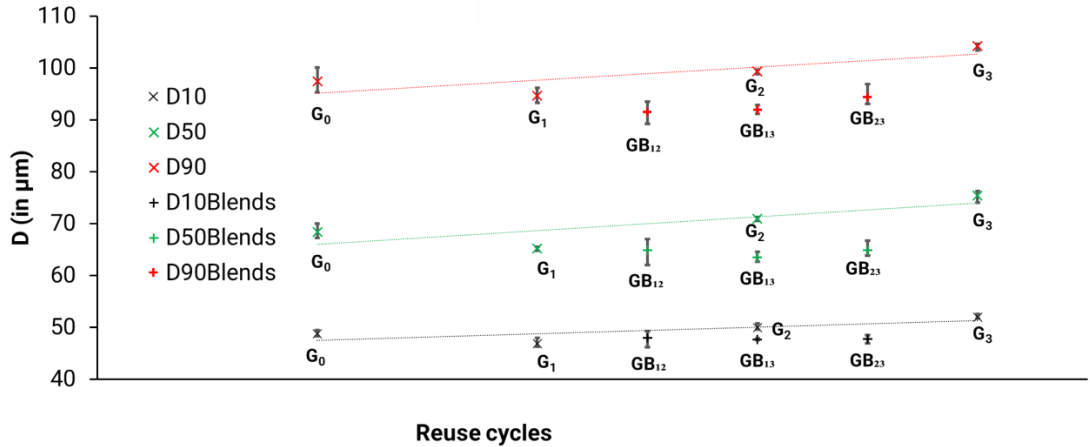

Figure 3 $D_{10}$, $D_{50}$ and $D_{90}$ values; for all powder types. Error bars in represent the min-max range.

The bulk density ($\rho_0$) measured with the GranuPack instrument indicate that the values for the individual genesis powders and the powder blends do not significantly change with reuse (as seen in Figure 4(a)). Similar results were observed by Tang et al. [18] where the $\rho_0$ remained unchanged after 21 reuse cycles. A powder with good flowability is usually characterized by a high $\rho_0$ value. This is because free flowing particles (with minimum interparticle adhesion) would be able to find an optimum arrangement and pack densely, therefore corresponding to a higher $\rho_0$ value. For such a powder, the possibility of a density increase during tapping is limited and therefore the ratio of tap density ($\rho_{500}$) to $\rho_0$ (also known as the Hausner ratio) would be close to unity.

The $\rho_{500}$ values are presented in Figure 4(a). It is observed that the tap density increases with the number of reuse cycles. For the most part, the tap density values for the powder blends lie between their respective genesis powders. The difference between the bulk and tapped densities is significant and increases with an increase in number of reuse cycles. Interparticulate interactions are usually larger for poorly flowing powders and therefore a greater difference between the bulk and tapped densities is observed [27]. This suggests that the powder flowability is reduced when the number of reuse cycles increases.

The Carr index ($C$) is the measure of the extent to which a powder can be compressed (without deforming the particles). The compressibility of the powder is expected to affect the continuity and uniformity of the powder layer, with lower $C$ values in favor of the formation of denser layers. As mentioned earlier, $H_r$ is an index that helps assess the flowability of the powder. According to Goyal et al. [30], for excellent compressibility and flowability, the $C$ (%) and , $H_r$ should be lower than 10 % and 1.11, respectively. The $H_r$ and $C$ values measured in this study, are presented in Figure 4(b)). When looking at all powder types, a good correlation is observed between the $H_r$ values and number of reuse cycles as well as Carr index and number of reuse cycles such that an increase in these metrics is observed with an increase in number of reuse cycles. For the most part, the $H_r$ and $C$ values for the powder blends lie between their respective genesis powders. All the values indicate that the flowability and compressibility is excellent (as defined by Goyal et al. [30]), however, the increasing trend suggests a degradation in the powder flowability characteristics. These plots strongly indicate that the powder has deteriorated from its virgin state. This observation is also supported by the general increase of PSD and changes in powder morphology observed as a function of increased reuse cycles.

      This degradation of flow properties of the powder blends, as well as the reused individual genesis powders is attributed to the deviation in the powder morphology as observed in the SEM micrographs (Figure 2). Such deviations from the spherical morphology are expected to not only decrease flowability but may also lead to uneven layer formation during raking and ultimately may result in powder bed non-uniformity across the build bed as mentioned in Table 1. On the other hand, the virgin (or $G_0$) powder is observed to be more spherical and therefore flows easily due to lower surface friction and mechanical interlocking, thus displaying a lower value for $H_r$ and $C$ metrics. The packing ability of powder particles influences the sintering of the powder layer [31]. As Neira-Arce [32] describes, uniform and homogeneous layers are crucial to ensure that there is proper heat conduction and for achieving dimensional accuracy, which in turn reduces the risk of swelling or overheating in EB-PBF parts.

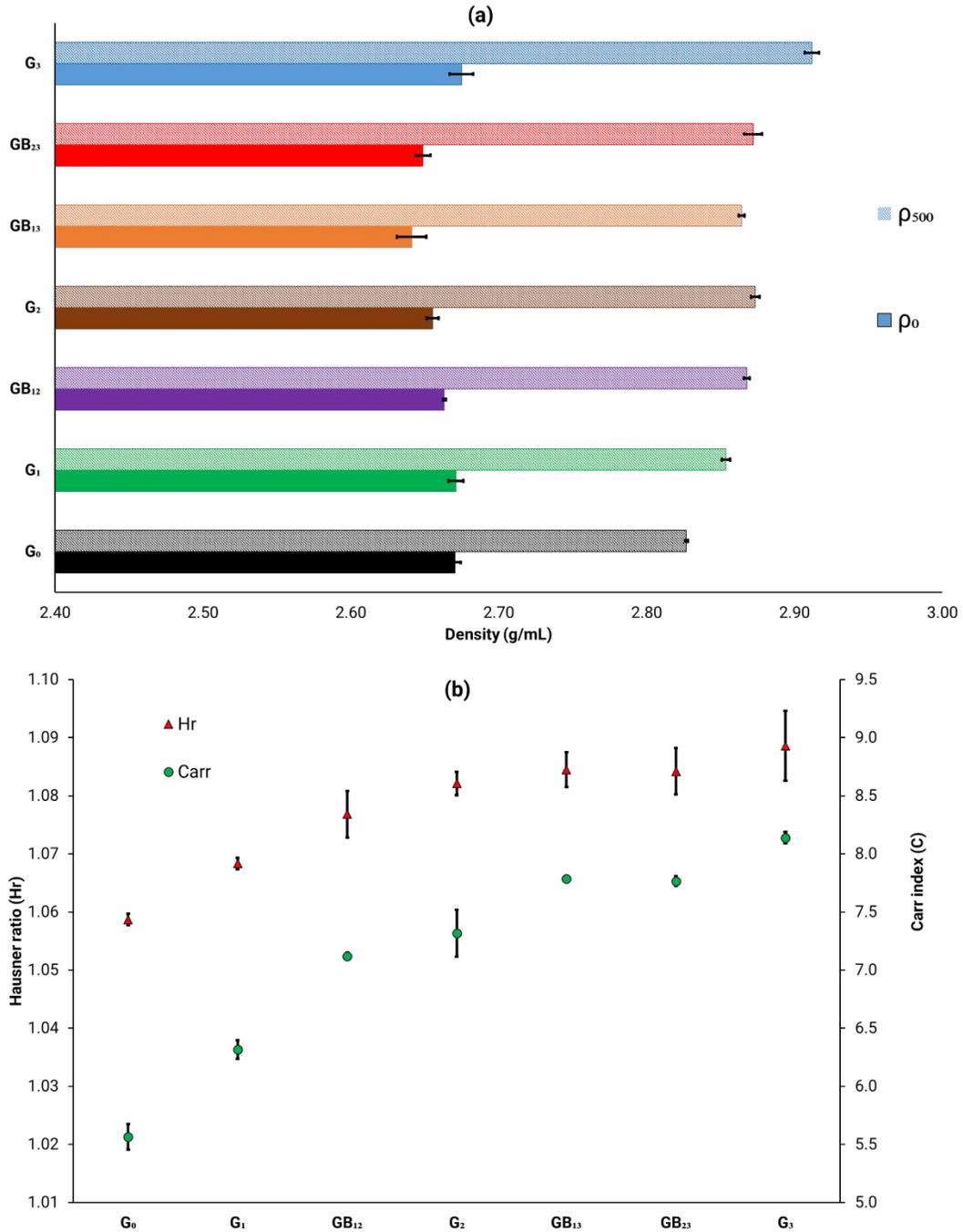

Figure 4 (a) $\rho_0$ and $\rho_{500}$ values (b) Hr and C values; for all powders. Error bars represent the standard deviation.

The BFE and SE (Figure 5(b)) values for the reused powders and blends have also increased with an increase in number of reuse cycles. The effect of larger BFE and smaller bulk density of the powder blends might result in a more uneven layer distribution. When looking at the individual genesis powders (Figure 5(b)), it can be concluded that both BFE and SE show an increase in values with an increase in number of reuse cycles, with $G_3$ being an exception. Similarly, there is a good correlation between BFE and SE (Figure 5(a)), for all powder types, where an increase in BFE leads to an increase in the SE values.

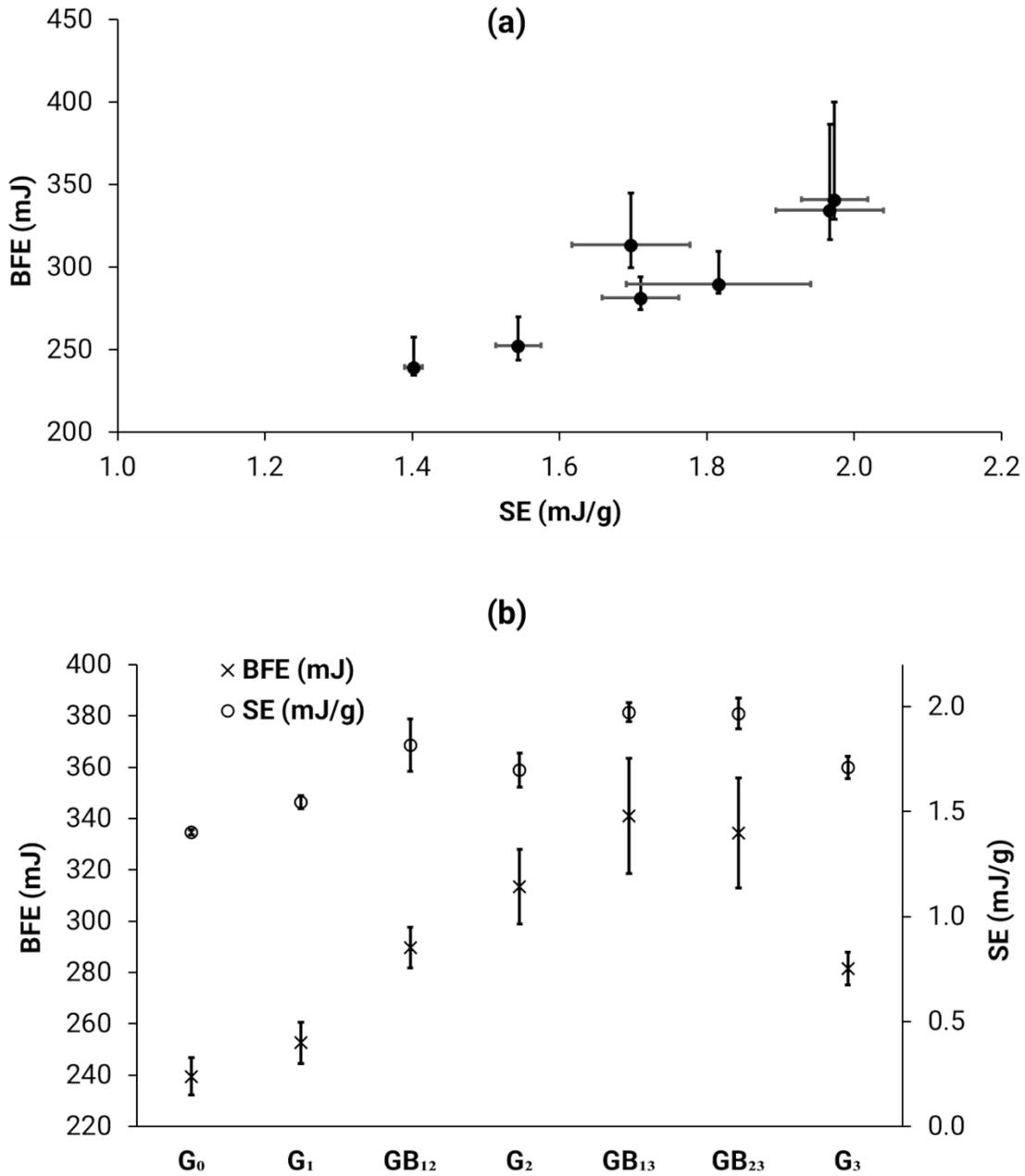

Figure 5 (a) BFE v/s SE for all powder types (b) BFE and SE values. Error bars represent the min-max range.

Strondl et al. [5] observed similar results in EB-PBF powders where the BFE and SE values increased after reusing. Clayton et al. [33] also compared the BFE results for virgin and reused powders and concluded that processing significantly increases the BFE values for reused powders. Both BFE and SE results suggest that the reused powders and blends would not flow as freely as $G_0$. The energy to move the blade in the rheometer is increasing for the reused powders and blends mainly due to their deviation from an overall spherical shape (affected by the pre-heating, blasting and sieving procedures) which causes greater interparticle interactions during testing of the powder samples in the FT4 instrument. As explained by Clayton et al. [33], a higher SE indicates increased mechanical interlocking and friction between particles that may potentially lead to flow problems. An interesting observation is that the BFE and SE values for the $G_3$ powder is lower than the $G_2$ powder. The reason behind this behavior is not well understood and therefore this powder type needs to be studied more extensively. but the authors speculate that the blends have a much higher concentration of non-spherical particles when compared to their respective genesis powders.

Figure 6(a) presents a plot of the dynamic angle of repose v/s rotating drum speed for all powder types. It is observed that below 25 rpm, it is difficult to make a differentiation between the various powders and all powder have excellent flowability at 2 rpm. Above 25 rpm, the AOR decreases with an increase in number of reuse cycles. As an example, from Figure 6(c), it can be observed that at the highest speed (i.e., 60 rpm), the AOR decreases from $G_1$ to $G_3$. It should be noted that $G_0$ stands out, probably because this powder has not undergone any processing in the machine (pre-heating, blasting, sieving).

The AOR values at the lowest and highest drum rotation speed (i.e., 2 rpm and 60 rpm, respectively) were correlated with the particle size $D_{10}$, $D_{50}$ and $D_{90}$ values. As can be seen from Figure 6(e) and (f), an excellent correlation is observed where an increase in particle size leads to a decrease in the AOR value. It is recognised that larger particles tend to flow more easily than finer powder. It has been observed that the $D_{10}$, $D_{50}$ and $D_{90}$ values increase with an increase in number of reuse cycles. Therefore, this increase in particle size is leading to a better flow in the rotating drum, thus suggesting that the flowability of the reused powders is better in this instrument.

Figure 6(b) presents the Cohesive Index (CI) at various drum rotation speeds for all powder types. The CI metric is linked to the fluctuations of the interface between the powder and air. The dynamic cohesive index of a powder depends on the magnitude of the cohesive forces between the particles. Therefore, a value closer to zero corresponds to a non-cohesive powder. Per the flow guidelines provided in the GranuDrum instrument software, threshold values for cohesive index are: < 5 very good, 5-10 good, 10-20 fair, 20-30 passable, 30-40 poor, > 40 very poor. The CI follows the same trend (as observed in Figure 6(b) and (d)) as the AOR curves, where an increase in number of reuse cycles lead to a decrease in the CI value (except $G_0$). An interesting observation is the fact that the CI values increase with an increase in number of reuse cycles (except $G_0$).

The GranuDrum data interpretation guide mentions that the flowability of a powder is measured as a function of the shearing rates, and therefore rheological properties like shear thinning or shear thickening could be evaluated with this instrument. If the powder AOR and CI increase with an increase in drum speed, the powder is said to show a shear-thickening behavior. All powders in this study show a shear-thickening behavior (Figure 6(a) and (b)). A powder material that shows a constant shear-thickening behavior is known to be a poor candidate for a dynamic process. That being said, there are portions in the CI v/s drum rotation speed curve Figure 6(b) where a somewhat constant cohesive index can be observed between speeds of 15 rpm and 40 rpm. The GranuDrum instrument helps identify an optimum raking/re-coating speed. The relationship between the drum rotating speed and surface flow speed (i.e., raking speed in mm/s) is displayed in Figure 6(g). It has been mentioned by the machine manufacturer [34], [35], that when selecting raking speeds, one should look at areas that display a constant cohesive index. As mentioned earlier, drum rotating speeds between 15 rpm and 40 rpm (corresponding to 75 mm/s and 175 mm/s respectively, according to Figure 6(g)) show a constant cohesive index. Therefore, raking speeds between 75 mm/s and 175 mm/s should be considered when using powders for EB-PBF processes.

      The effect of powder reused on interparticle cohesion is contrary to the measurements done with FT4 instrument (higher BFE and SE with reuse). It is not clear at the moment whether the increase of the PSD with reuse is only associated with the agglomeration, the sieving, or a combination of both. Difference in flow regime and the sensitivity of the flowability to multiple powder modifications may explain the discrepancy in these results.

      As mentioned by Brika et al. [36], powder flowability is not an inherent material property and is the ability of the powder to flow in a desired manner in a particular instrument. A powder may perform well in a certain instrument/piece of equipment while may perform poorly in another. Thus, additional tests are required to better understand which regime better represents the behaviour of the powder in an additive manufacturing (AM) machine.

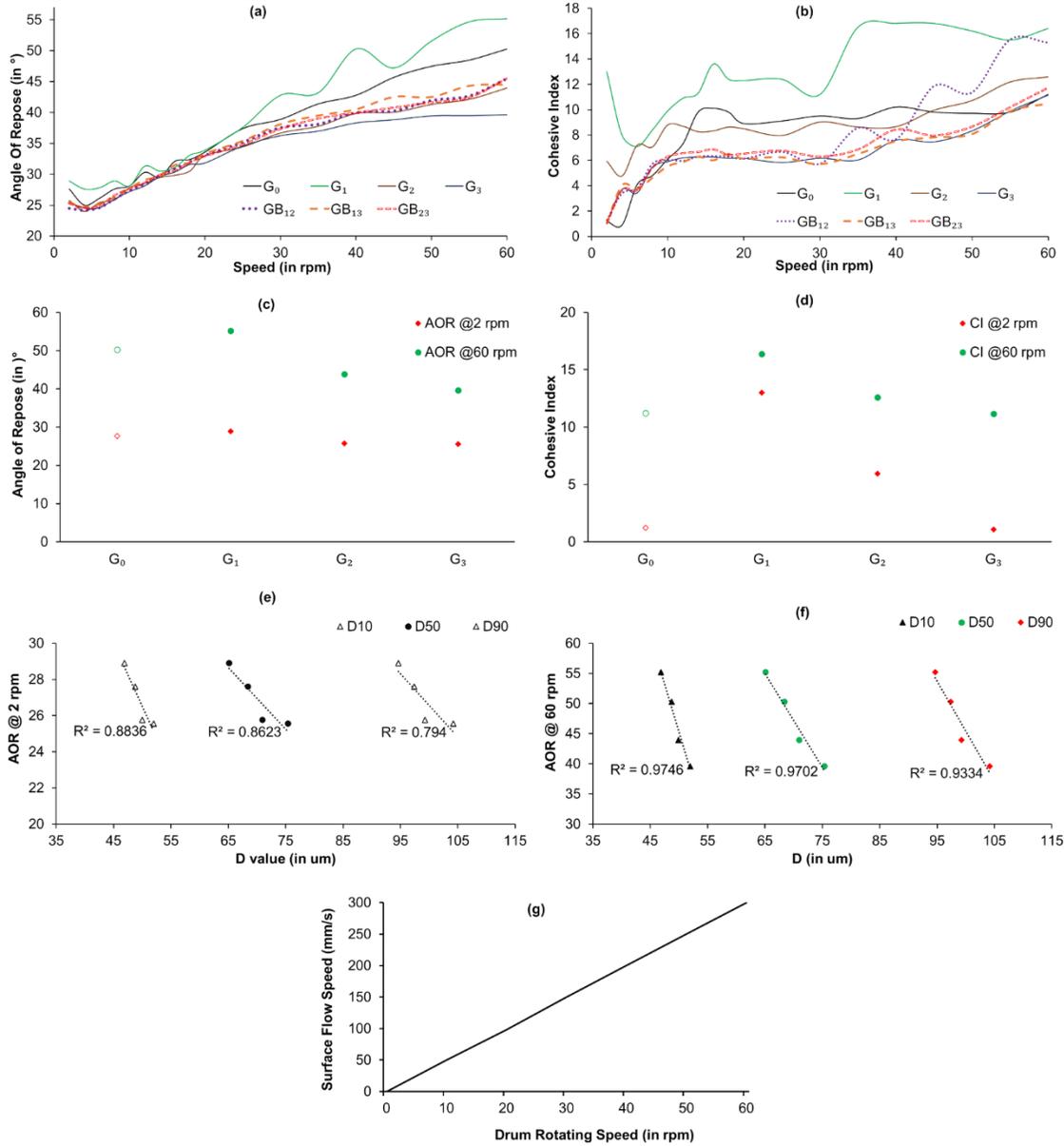

Figure 6 (a) AOR v/s speed of drum rotation for all powders (b) CI v/s speed of drum rotation for all powder (c) AOR at 2 and 60 rpm v/s number of reuse cycles (d) CI at 2 and 60 rpm v/s number of reuse cycles (e) AOR at 2 rpm v/s particle size (D values) for individual genesis powders (f) AOR at 60 rpm v/s particle size (D values) for individual genesis (g) Relationship between drum rotating speed and surface flow speed as identified by Granutools [34]. Note than $G_0$ powder is considered an anomaly due to the fact that it has not undergone any processing, blasting, or sieving procedure.

Figure 7 (a) and (b) shows the average values for oxygen and nitrogen concentration increases with number of reuse cycles. Specifically, a 37% and 44% increase was observed in the oxygen and nitrogen concentration, respectively, for $G_3$ powder when compared to the $G_0$ powder. However, the oxygen remains lower than 0.18 wt.% and below the limit outlined by ASTM F2924-14 [11]. A logarithmic trendline seems to better fit the results when compared to the linear trendline (Figure 7(a)). From the trendline, it can be deduced that the $O_2$ concentration will exceed the 0.2 wt.% limit by 5-6 reuse cycles. The nitrogen concentration remains between 0.016 wt.% and 0.023 wt.% (Figure 7(b)) which is well below the 0.05 wt.% limit outlined by ASTM 2924-14 [11]. Increase of nitrogen concentration with the number of reuse cycles is observed; the logarithmic trend line suggest that this nitrogen pick-up may reach a saturation level, suggesting that this is a surface contamination (formation of nitride or local concentration at the surface). The trend suggests that the maximum (i.e., 0.05 wt.%) will not be reached before a large number of reuse cycles.

Some other studies have also looked at the increase in oxygen and nitrogen content with increase in number of reuse cycles. Ghods et al. [37] showed that the oxygen content was > 0.20 wt.% after 11 reuse cycles. Grainger [29] observed a linear increase in oxygen and nitrogen concentrations in Grade 23 Ti-6Al-4V powder after reusing in the laser PBF process. However, after 30 builds, the oxygen concentration remained below 0.20 wt.%. These values are attributed to the fact that the starting oxygen concentration in Grade 23 Ti-6Al-4V is much lower than that of Grade 5 Ti-6Al-4V. Nandwana et al. [38] observed an increase in the oxygen concentration from 0.13 wt.% to 0.18 wt.%, however their nitrogen concentration remained the same over 5 reuse cycles for a Ti-6Al-4V powder.

There are numerous factors which contribute to oxygen pick-up in Ti-6Al-4V powders processed through EB-PBF. One of the possible reasons for the increase of oxygen pickup is the fact that the powder is exposed to the ambient atmosphere and moisture when transferred from the machine to the PRS for part recovery, and then transferred from the PRS to the sieve. A study performed by Vluttert [39] shows that Ti-6Al-4V can pick up about 0.2 wt. % moisture (relative to dry weight) when left in an environment that was set to 25 °C and 90% humidity. Therefore, the presence of water vapor in the air may have a reasonable impact on oxygen pickup. Montelione et al. [37] compare the EB-PBF process to gas tungsten arc welding and suggest that when the powder is exposed to air, the water molecules adsorbed on the particle surface may dissociate during heating, allowing the oxygen to diffuse in the metal alloy. In addition, the deformation caused by blasting and sieving can also lead to an acceleration of the oxidation due to creation of new oxidation prone surfaces. Shvab et al. [40] describe that this will lead to a passive oxide layer formation that may diffuse inward and reform during heating cycles. Furthermore, oxygen pickup can also occur if the machine is not under vacuum when the powder is stored in the hoppers in between the builds. Finally, the major source of oxygen pickup is the reaction of titanium at high temperature with residual oxygen in the atmosphere of the machine. In addition, Mizuno et al. [41] noted that all oxides in titanium alloys dissolve above 400 °C and the oxygen diffuses into the bulk through the bulk diffusion process. The EB-PBF process takes place at a relatively high temperature of 730°C where the titanium can react with residual oxygen, while the high temperatures can result in oxygen diffusion as described by Attalla et al. [42] (build chamber ≈ $10^{-4}$ mbar and electron beam column ≈ $10^{-7}$ mbar). The oxygen and nitrogen can dissolve interstitially into the titanium lattice and form oxides and nitrides which can also impact the mechanical behavior. As stated by Donachie [43], an increase in the oxygen and nitrogen content in solution can lead to an increase in the strength, and decrease in the ductility which further leads to embrittlement.

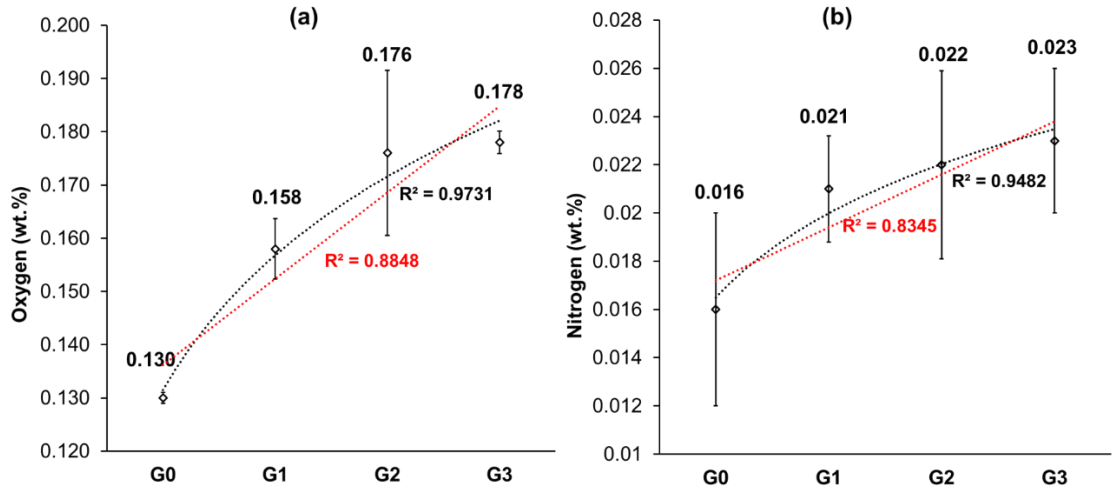

Figure 7 (a) Oxygen and (b) Nitrogen concentration for $G_0$, $G_1$, $G_2$, and $G_3$ powder types along with linear (in red) and logarithmic (in black) trendlines. Error bars represent the standard deviation.

### 3.2 Evaluation of the relative performance of powders

It is quite challenging to deploy a unified metric to capture the absolute suitability of powders for PBF AM processes. The powder metrics measured are very different in nature and may impact the PBF process and final part properties in different ways. For instance, powder metrics variations may have opposite effects, i.e., some characteristics may lead to an improved behavior in the AM process, others may lead to a deterioration of the AM process behaviour, while some powder metrics may have limited effect. In literature, this is poorly understood at the moment. Variability is still relatively high for some of the powder measurement techniques, and are influenced by the testing equipment type, the testing conditions, the powder storage and handling conditions, the operator skill, the calibration of instruments, and many other factors. Care must be exercised in interpreting each powder metric result individually. Although it is challenging to capture the absolute suitability of powders for PBF AM processes via a consolidated index, there is a potential to describe the relative change in powder properties with respect to a reference state, if such reference should exist. A reference state, for instance, is often considered to be the virgin powder. Two examples of calculations of relative changes from a reference state are presented below.

#### 3.2.1 Relative Performance Evaluated via Radar Diagram

To understand the sensitivity of response to change of the various powder characteristics, from a reference powder ($G_0$ in this work), the various properties measured were normalized to a range of 0 to 1 and a radar diagram of the normalized indices for the $G_0$, $G_1$, $G_2$, and $G_3$ powders was plotted in Figure 8. The normalization was done to transform the data into a dimensionless data sequence for ease of comparability. For this study, the approach used for normalizing all metrics was taken from Mehat et al. [44]:

$$x_i^*(k) = \frac{x_i^{(0)}(k) - \min x_i^{(0)}(k)}{\max x_i^{(0)}(k) - \min x_i^{(0)}(k)}$$

where $x_i^{(0)}(k)$ is the measurement of the quality characteristic, $\max x_i^{(0)}(k)$ is the largest value of $x_i^{(0)}(k)$, and $\min x_i^{(0)}(k)$ is the smallest value of $x_i^{(0)}(k)$.

In accordance with the powder suitability criteria created in Table 1, the smaller the area covered by the radar diagram, the higher the powder suitability for the EB-PBF process. It is observed that the $G_0$ powder has the smallest area, followed by $G_1$, $G_2$ and $G_3$, indicating that the powder is becoming less-suitable with increasing number of reuse cycles.

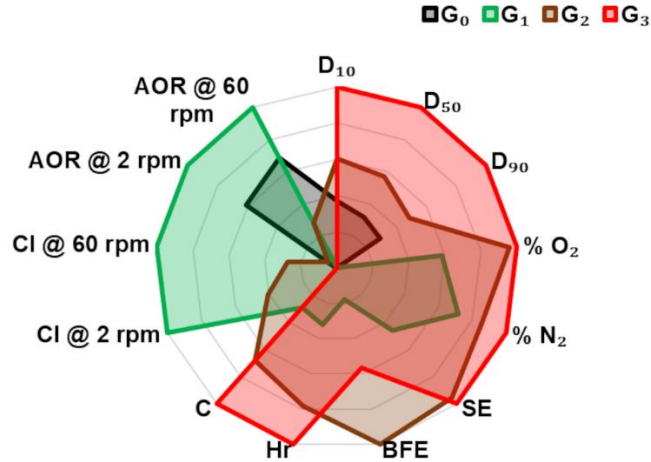

Figure 8 Radar diagram comparing the performance of $G_0$, $G_1$, $G_2$, and $G_3$ powders w.r.t the D values, SE, BFE, Nitrogen (in wt.%), Oxygen (in wt.%), $\rho_0$, Hr, C, CI, and AOR.

### 3.2.2 Relative Performance Evaluated via Performance Index

Efforts in assessing powder suitability for powder bed fusion were undertaken by Brika et al. [36]; they differentiated powders and assessed three powder types by calculating a figure of merit called AMS (Additive Manufacturing Suitability). While in this present study, the authors are not assessing an absolute suitability criteria due to challenges aforementioned, this approach is deployed herein to assess if this index can be used to quantify the relative changes in powder properties with respect to a reference state ($G_0$ in this work). Since the current study focuses on EB-PBF powder suitability, the index will be referred as "ESF" (Electron-beam powder bed fusion Suitability Factor) and is inspired from Brika et al. [36]. As such, for each powder reuse cycle, a cumulative sum of the normalized index presented in Figure 8 will provide an objective function:

$$\text{ESF} = \frac{(D10 + D50 + D90 + O2 + N2 + SE + BFE + H_r + C + CI @ 2 \text{ rpm} + CI @ 60 \text{ rpm} + AOR @ 2 \text{ rpm} + AOR @ 60 \text{ rpm})}{13}$$

The ESF values for $G_0$, $G_1$, $G_2$, and $G_3$ powders are 0.176, 0.504, 0.609, and 0.659, respectively. It is assumed that the lower deviation from the ESF value with respect to $G_0$, the more the properties of the powder will depart from the reference state. This is another example of efforts towards expressing the relative change in powder characteristics.

As a conclusion, this study looked into comparing the influence of multiple reuse cycles, as well as powder blends created from reused powder, through various performance metrics such as morphology, size distribution, basic flow energy, specific energy, bulk density and tap density after 500 taps, Hausner ratio and Carr index, cohesive index, angle of repose, oxygen content, and nitrogen content. In accordance with the authors' hypotheses established in Shanbhag and Vlasea [23], it was observed that reusing modifies the powder significantly when assessing the various performance metrics. Future research efforts will rely on this present study and will look into assessing the parts that were printed in the powder re-use campaigns to be able to understand the impact of reusing powders on part performance including surface roughness, porosity, and powder cake density.

# 4 Conclusions

Investigation into the effect of Plasma atomized Grade 5 Ti-6Al-4V powder reuse on the powder properties as well as properties of powder blends led to the following conclusions:

(1) The SEM micrographs of the various powder types show extensive physical modification to the surface of the particles, with increasing degree of powder reuse. The micrographs depict features such as elongated particles, broken particles, deformed particles, particle with molten specks and agglomerates. The broke, shattered, clip-clap and deformed particles are attributed to the powder recovery process (i.e., blasting, sieving) and tumbling process; and the particles with molten specks, elongated particles and agglomerated particles are attributed to the high temperature conditions leading to overheating and smelting of particles and satellites.

(2) The $D_{10}$, $D_{50}$ and $D_{90}$ values increase with increasing degree of powder reuse. This observation is attributed to the agglomeration of powder particles.

(3) The $\rho_0$ remained unchanged for all powders; however, the $\rho_{500}$ increases with increasing degree of powder reuse. The $H_r$ and $C$ values show an increase with an increase in number of reuse cycles. This trend indicates modification of reused powders from their virgin state. These observations have been attributed to the deviations from spherical morphology, for the reused powder, as observed in the SEM micrographs. Due to these deviations, uneven raking and non-homogenous layers may be obtained.

(4) The BFE and SE measured using powder rheology increase with increasing degree of powder reuse. This suggests that the reused powder is more cohesive than the virgin powder. This behavior is attributed to the mechanical interlocking and friction between particles (caused by the non-spherical morphology).

(5) The dynamic AOR and CI values increase with an increase in number of reuse cycles. This behavior is attributed to the agglomeration and increase of the $D_{10}$, $D_{50}$ and $D_{90}$ with the number of reused cycles. The variations of the metrics with drum rotating speed may help identify a range of optimum raking speeds that can be used for these powders when being used in the EB-PBF machine.

(6) The $O_2$ and $N_2$ concentration remain below the limits outlined by ASTM F2924. However, a gradual increase has been observed with increasing degree of powder reuse. Specifically, a 37% and 44% increase was observed in the oxygen and nitrogen concentration. From the trendline, it can be deduced that the $O_2$ concentration will exceed the 0.2 wt.% limit by 5-6 reuse cycles.

(7) A unified powder quality score or powder quality metric was established to compare the effect of powder reuse on the various powder performance metrics.

## 5 Funding

The authors appreciate the partial funding support received from the Natural Sciences and Engineering Research Council (NSERC) Holistic Innovation in Additive Manufacturing (HI-AM) grant NETGP 494158 - 16. The authors also appreciate the research contributions and funding support from the National Research Council, Canada (NRC) under the "Additive Manufacturing Materials and Processes" collaboration program 080531, sub-project SUB03280.

## 6 Acknowledgements

The authors would like to acknowledge Dr. Louis-Philippe Lefebvre, Research Officer at National Research Council Canada, for his significant efforts in the review of the paper, in providing editorial comments, scientific and technical feedback. In addition, the authors would like to acknowledge the support from the Multi Scale Additive Manufacturing Lab (MSAM, University of Waterloo) technical staff Jerry Ratthapakdee for assisting with the manufacturing process and interns (specifically Kate Pearson and Cecilia Cancellara) for assisting with powder characterization datasets, as well as Shirley Mercier from NRC for conducting the LECO analysis.

## 7 Author Contributions

Conceptualization, G.S. and M.V.; investigation, G.S.; methodology, G.S.; project administration, G.S.; supervision, M.V.; writing — original draft, G.S..; writing — review and editing, M.V.; funding acquisition, M.V.